\documentclass[12pt,twoside,english]{article}
\usepackage[T1]{fontenc}
\usepackage[latin9]{inputenc}
\usepackage{geometry}
\usepackage{color}
\usepackage{babel}
\usepackage{amstext}
\usepackage{amssymb}
\usepackage[unicode=true,
 bookmarks=true,bookmarksnumbered=true,bookmarksopen=true,bookmarksopenlevel=1,
 breaklinks=false,pdfborder={0 0 1},backref=false,colorlinks=true]
 {hyperref}
\hypersetup{pdftitle={The LyX Tutorial},
 pdfauthor={LyX Team},
 pdfsubject={LyX-documentation Tutorial},
 pdfkeywords={LyX, documentation},
 linkcolor=black, citecolor=black, urlcolor=blue, filecolor=blue,pdfpagelayout=OneColumn, pdfnewwindow=true, pdfstartview=XYZ, plainpages=false}
\usepackage{breakurl}

\makeatletter

\DeclareRobustCommand{\greektext}{%
  \fontencoding{LGR}\selectfont\def\encodingdefault{LGR}}
\DeclareRobustCommand{\textgreek}[1]{\leavevmode{\greektext #1}}
\DeclareFontEncoding{LGR}{}{}
\DeclareTextSymbol{\~}{LGR}{126}
\newcommand{\lyxmathsym}[1]{\ifmmode\begingroup\def\b@ld{bold}
  \text{\ifx\math@version\b@ld\bfseries\fi#1}\endgroup\else#1\fi}

\usepackage{ifpdf} 
\ifpdf 

 \IfFileExists{lmodern.sty}{\usepackage{lmodern}}{}

\fi 

\let\myTOC\tableofcontents
\renewcommand\tableofcontents{%
  \frontmatter
  \pdfbookmark[1]{\contentsname}{}
  \myTOC
  \mainmatter }

\def\LyX{\texorpdfstring{%
  L\kern-.1667em\lower.25em\hbox{Y}\kern-.125emX\@}
  {LyX}}

\makeatother

\begin{document}

\title{{\Large RELATIONAL QUANTUM MECHANICS}%
\thanks{The present paper will appear as individual chapter in the book \textquotedbl{}Quantum
Mechanics\textquotedbl{}, to be published by InTech.%
}{\Large{} }}

\author{{\Large A. Nicolaidis}\medskip \\Theoretical Physics Department\\Aristotle
University of Thessaloniki\\54124 Thessaloniki, Greece\\nicolaid@auth.gr}
\maketitle
\begin{abstract}
We suggest that the inner syntax of Quantum Mechanics is relational
logic, a form of logic developed by C. S. Peirce during the years
1870 \textendash{} 1880. The Peircean logic has the structure of category
theory, with relation serving as an arrow (or morphism). At the core
of the relational logical system is the law of composition of relations.
This law leads to the fundamental quantum rule of probability as the
square of an amplitude. Our study of a simple discrete model, extended
to the continuum, indicates that a finite number of degrees of freedom
can live in phase space. This \textquotedblleft{}granularity\textquotedblright{}
of phase space is determined by Planck\textquoteright{}s constant
h. We indicate also the broader philosophical ramifications of a relational
quantum mechanics. 
\end{abstract}

\section*{{\large Introduction}}

\qquad{}Quantum mechanics (QM) stands out as the theory of the 20th
century, shaping the most diverse phenomena, from subatomic physics
to cosmology. All quantum predictions have been crowned with full
success and utmost accuracy. Yet, the admiration we feel towards QM
is mixed with surprise and uneasiness. QM defies common sense and
common logic. Various paradoxes, including Schrodinger\textquoteright{}s
cat and EPR paradox, exemplify the lurking conflict. The reality of
the problem is confirmed by the Bell\textquoteright{}s inequalities
and the GHZ equalities. We are thus led to revisit a number of old
interlocked oppositions: operator \textendash{} operand, discrete
\textendash{} continuous, finite \textendash{}infinite, hardware \textendash{}
software, local \textendash{} global, particular \textendash{} universal,
syntax \textendash{} semantics, ontological \textendash{} epistemological. 

\qquad{}The logic of a physical theory reflects the structure of
the propositions describing the physical system under study. The propositional
logic of classical mechanics is Boolean logic, which is based on set
theory. A set theory is deprived of any structure, being a plurality
of structure-less individuals, qualified only by membership (or non-membership).
Accordingly a set-theoretic enterprise is analytic, atomistic, arithmetic.
It was noticed as early as 1936 by Neumann and Birkhoff that the quantum
real needs a non-Boolean logical structure. On numerous cases the
need for a novel system of logical syntax is evident. Quantum measurement
bypasses the old disjunctions subject-object, observer-observed. The
observer affects the system under observation and the borderline between
ontological and epistemological is blurred. Correlations are not anymore
local and a quantum system embodies multiple entanglements. The particular-universal
dichotomy is also under revision. While a single quantum event is
particular, a plethora of quantum events leads to universal patterns.
Viewing the quantum system as a system encoding information, we understand
that the usual distinction between hardware and software is not relevant.
Most importantly, if we consider the opposing terms being-becoming,
we realize that the emphasis is sifted to the becoming, the movement,
the process. The underlying dynamics is governed by relational principles
and we have suggested \cite{key-1} that the relational logic of C.
S. Peirce may serve as the conceptual foundation of QM. 

\qquad{}Peirce, the founder of American pragmatism, made important
contributions in science, philosophy, semeiotics and notably in logic.
Many scholars (Clifford, Schröder, Whitehead, Lukasiewicz) rank Peirce
with Leibniz and Aristotle in the history of thought. Logic, in its
most general sense, is the formal science of representation, coextensive
with semeiotics. Algebraic logic attempts to express the laws of thought
in the form of mathematical equations, and Peirce incorporated a theory
of relations into algebraic logic \cite{key-2,key-3}. Relation is
the primary irreducible datum and everything is expressed in terms
of relations. A relational formulation is bound to be synthetic, holistic,
geometric. Peirce invented also a notation for quantifiers and developed
quantification theory, thus he is regarded as one of the principal
founders of modern logic.

\qquad{}In the next section we present the structures of the relational
logic and a representation of relation which will lead us to the probability
rule of QM. In the third section we analyze a discrete system and
demonstrate the non-commutation of conjugate operators. In the last
section we present the conclusions and indicate directions for future
work.

\section*{{\large The logic of relations and the quantum rules}}

\qquad{}The starting point is the binary relation $S_{i}RS_{j}$
between the two \textquoteright{}individual terms\textquoteright{}
(subjects) $S_{j}$ and $S_{i}$. In a short hand notation we represent
this relation by $R_{ij}$. Relations may be composed: whenever we
have relations of the form $R_{ij}$, $R_{jl}$, a third transitive
relation $R_{il}$ emerges following the rule \cite{key-2,key-3}
\begin{equation}
R_{ij}R_{kl}=\lyxmathsym{\textgreek{d}}_{jk}R_{il}
\end{equation}
In ordinary logic the individual subject is the starting point and
it is defi{}ned as a member of a set. Peirce, in an original move,
considered the individual as the aggregate of all its relations 
\begin{equation}
S_{i}=\sum_{j}R_{ij}.
\end{equation}
It is easy to verify that the individual $S_{i}$ thus defi{}ned is
an eigenstate of the $R_{ii}$ relation
\begin{equation}
R_{ii}S_{i}=S_{i}.
\end{equation}
The relations $R_{ii}$ are idempotent
\begin{equation}
R_{ii}^{2}=R_{ii}
\end{equation}
and they span the identity 
\begin{equation}
\sum_{i}R_{ii}=\mathbf{1}
\end{equation}
The Peircean logical structure bears great resemblance to category
theory, a remarkably rich branch of mathematics developed by Eilenberg
and Maclane in 1945 \cite{key-4}. In categories the concept of transformation
(transition, map, morphism or arrow) enjoys an autonomous, primary
and irreducible role. A category \cite{key-5} consists of objects
A, B, C,... and arrows (morphisms) f, g, h,... . Each arrow f is assigned
an object $A$ as domain and an object $B$ as codomain, indicated
by writing $f:A\rightarrow B$. If $g$ is an arrow $g:B\rightarrow C$
with domain $B$, the codomain of f, then f and g can be \textquotedblleft{}composed\textquotedblright{}
to give an arrow $gof:A\rightarrow C$. The composition obeys the
associative law $ho(gof)=(hog)of$. For each object $A$ there is
an arrow $1_{A}:A\rightarrow A$ called the identity arrow of $A$.
The analogy with the relational logic of Peirce is evident, $R_{ij}$
stands as an arrow, the composition rule is manifested in eq. (1)
and the identity arrow for $A\equiv S_{i}$ is $R_{ii}$. There is
an important literature on possible ways the category notions can
be applied to physics; specifi{}cally to quantising space-time \cite{key-6},
attaching a formal language to a physical system \cite{key-7}, studying
topological quantum fi{}eld theories \cite{key-8,key-9}, exploring
quantum issues and quantum information theory \cite{key-10}. 

\qquad{}A relation $R_{ij}$ may receive multiple interpretations:
as the proof of the logical proposition $i$ starting from the logical
premise $j$, as a transition from the $j$ state to the $i$ state,
as a measurement process that rejects all impinging systems except
those in the state $j$ and permits only systems in the state $i$
to emerge from the apparatus. We proceed to a representation of $R_{ij}$
\begin{equation}
R_{ij}=\left|r_{i}\right\rangle \left\langle r_{j}\right|
\end{equation}
where state $\left\langle r_{i}\right|$ is the dual of the state$\left|r_{i}\right\rangle $
and they obey the orthonormal condition 
\begin{equation}
\left\langle r_{i}\right|\left.r_{j}\right\rangle =\delta_{ij}
\end{equation}
It is immediately seen that our representation satisfi{}es the composition
rule eq. (1). The completeness, eq.(5), takes the form 
\begin{equation}
\sum_{i}\left|r_{i}\right\rangle \left\langle r_{i}\right|=\mathbf{1}
\end{equation}
All relations remain satisfi{}ed if we replace the state $\left|r_{i}\right\rangle $
by $\left|\varrho_{i}\right\rangle $, where
\begin{equation}
\left|\varrho_{i}\right\rangle =\frac{1}{\sqrt{N}}\sum_{n}\left|r_{i}\right\rangle \left\langle r_{n}\right|
\end{equation}
with $N$ the number of states. Thus we verify Peirce\textquoteright{}s
suggestion, eq. (2), and the state $\left|r_{i}\right\rangle $ is
derived as the sum of all its interactions with the other states.
$R_{ij}$ acts as a projection, transferring from one $r$ state to
another $r$ state 
\begin{equation}
R_{ij}\left|r_{k}\right\rangle =\delta_{jk}\left|r_{i}\right\rangle .
\end{equation}
We may think also of another property characterizing our states and
defi{}ne a corresponding operator
\begin{equation}
Q_{ij}=\left|q_{i}\right\rangle \left\langle q_{j}\right|
\end{equation}
with 
\begin{equation}
Q_{ij}\left|q_{k}\right\rangle =\delta_{jk}\left|q_{i}\right\rangle 
\end{equation}
and 
\begin{equation}
\sum_{i}\left|q_{i}\right\rangle \left\langle q_{i}\right|=\mathbf{1}.
\end{equation}
Successive measurements of the $q$-ness and $r$-ness of the states
is provided by the operator 
\begin{equation}
R_{ij}Q_{kl}=\left|r_{i}\right\rangle \left\langle r_{j}\right|\left.q_{k}\right\rangle \left\langle q_{l}\right|=\left\langle r_{j}\right|\left.q_{k}\right\rangle S_{il}
\end{equation}
with 
\begin{equation}
S_{il}=\left|r_{i}\right\rangle \left\langle q_{l}\right|.
\end{equation}
Considering the matrix elements of an operator $A$ as $A_{nm}=\left\langle r_{n}\right.\left|A\right|\left.r_{m}\right\rangle $
we fi{}nd for the trace 
\begin{equation}
Tr\left(S_{il}\right)=\sum_{n}\left\langle r_{n}\right.\left|S_{il}\right|\left.r_{n}\right\rangle =\left\langle q_{l}\right|\left.r_{i}\right\rangle .
\end{equation}
>From the above relation we deduce 
\begin{equation}
Tr\left(R_{ij}\right)=\delta_{ij}.
\end{equation}
Any operator can be expressed as a linear superposition of the $R_{ij}$
\begin{equation}
A=\sum_{i,j}A_{ij}R_{ij}
\end{equation}
with 
\begin{equation}
A_{ij}=Tr\left(AR_{ij}\right).
\end{equation}
The individual states can be redefi{}ned
\begin{eqnarray}
\left|r_{i}\right\rangle  & \rightarrow & e^{i\varphi_{i}}\left|r_{i}\right\rangle \\
\left|q_{i}\right\rangle  & \rightarrow & e^{i\theta_{i}}\left|q_{i}\right\rangle 
\end{eqnarray}
without aff{}ecting the corresponding composition laws. However the
overlap number $\left\langle r_{i}\right|\left.q_{j}\right\rangle $
changes and therefore we need an invariant formulation for the transition
$\left|r_{i}\right\rangle \rightarrow\left|q_{j}\right\rangle $.
This is provided by the trace of the closed operation $R_{ii}Q_{jj}R_{ii}$
\begin{equation}
Tr\left(R_{ii}Q_{jj}R_{ii}\right)\equiv p\left(q_{j},r_{i}\right)=\left|\left\langle r_{i}\right|\left.q_{j}\right\rangle \right|^{2}.
\end{equation}
The completeness relation, eq. (13), guarantees that $p\left(q_{j},r_{i}\right)$
may assume the role of a probability since 
\begin{equation}
\sum_{j}p\left(q_{j},r_{i}\right)=1.
\end{equation}
We discover that starting from the relational logic of Peirce we obtain
the essential law of Quantum Mechanics. Our derivation underlines
the outmost relational nature of Quantum Mechanics and goes in parallel
with the analysis of the quantum algebra of microscopic measurement
presented by Schwinger \cite{key-10-1}.

\section*{The emergence of Planck's constant}

\qquad{}Consider a chain of $N$ discrete states $\left|a_{k}\right\rangle $,
with $k=1,2,\ldots,N$. A relation $R$ acts like a shift operator
\begin{eqnarray}
R\left|a_{k}\right\rangle  & = & \left|a_{k+1}\right\rangle \\
R\left|a_{N}\right\rangle  & = & \left|a_{1}\right\rangle 
\end{eqnarray}
$N$ is the period of $R$
\begin{equation}
R^{N}=\mathbf{1}
\end{equation}
The numbers which satisfy $a^{N}=1$ are given by\textbf{
\begin{equation}
a_{k}=\exp\left(2\pi i\frac{k}{N}\right)\quad k=1,2,\ldots,N
\end{equation}
}Then we have 
\begin{equation}
R^{N}-1=\left(\frac{R}{a_{k}}\right)^{N}-1=\left[\left(\frac{R}{a_{k}}\right)-1\right]\sum_{j=0}^{N-1}\left(\frac{R}{a_{k}}\right)^{j}=0
\end{equation}
$R$ has a set of eigenfunctions
\begin{equation}
R\left|b_{i}\right\rangle =b_{i}\left|b_{i}\right\rangle 
\end{equation}
with $b_{i}$ the $N$-th root of unity $\left(b_{i}=a_{i}\right)$.
It is decomposed like
\begin{equation}
R=\sum_{j}b_{j}\left|b_{j}\right\rangle \left\langle b_{j}\right|
\end{equation}
Notice that we may write 
\begin{equation}
\left|b_{j}\right\rangle \left\langle b_{j}\right|=\frac{1}{N}\sum_{k=1}^{N}\left(\frac{R}{b_{j}}\right)^{k}
\end{equation}
 The above projection operator acting upon $\left|a_{N}\right\rangle $
will give
\begin{equation}
\left|b_{j}\right\rangle \left\langle b_{j}\right|\left.a_{N}\right\rangle =\frac{1}{N}\sum_{k=1}^{N}\left(\frac{1}{b_{j}}\right)^{k}\left|a_{k}\right\rangle \label{eq:32}
\end{equation}
 Matching from the right with $\left\langle a_{N}\right|$ we obtain
\begin{equation}
\left\langle a_{N}\right|\left.b_{j}\right\rangle \left\langle b_{j}\right|\left.a_{N}\right\rangle =\frac{1}{N}
\end{equation}
We adopt the positive root 
\begin{equation}
\left\langle b_{j}\right|\left.a_{N}\right\rangle =\frac{1}{\sqrt{N}}
\end{equation}
and equ. (\ref{eq:32}) becomes
\begin{equation}
\left|b_{j}\right\rangle =\frac{1}{\sqrt{N}}\sum_{k=1}^{N}\exp\left[-2\pi i\,\frac{jk}{N}\right]\left|a_{k}\right\rangle 
\end{equation}
Inversely we have the decomposition 
\begin{equation}
\left|a_{m}\right\rangle =\frac{1}{\sqrt{N}}\sum_{n=1}^{N}\exp\left[2\pi i\,\frac{mn}{N}\right]\left|a_{n}\right\rangle .
\end{equation}
We introduce another relation $Q$ acting like shift operator
\begin{eqnarray}
\left\langle b_{k}\right|Q & = & \left\langle b_{k+1}\right|\\
\left\langle b_{N}\right|Q & = & \left\langle b_{1}\right|
\end{eqnarray}
The relation $Q$ receives the decomposition 
\begin{equation}
Q=\sum_{j}a_{j}\left|a_{j}\right\rangle \left\langle a_{j}\right|
\end{equation}
Consider now 
\begin{equation}
\left\langle b_{k}\right|QR=\left\langle b_{k+1}\right|R=\exp\left[2\pi i\frac{\left(k+1\right)}{N}\right]\left\langle b_{k+1}\right|
\end{equation}
\begin{equation}
\left\langle b_{k}\right|RQ=\exp\left[2\pi i\frac{k}{N}\right]\left\langle b_{k}\right|Q=\exp\left[2\pi i\frac{k}{N}\right]\left\langle b_{k+1}\right|
\end{equation}
We conclude that the conjugate operators $R$ and $Q$ do not commute
\begin{equation}
QR=\exp\left[2\pi i\frac{1}{N}\right]RQ\label{eq:42}
\end{equation}
Similarly 
\begin{equation}
Q^{n}R^{m}=\exp\left[2\pi i\frac{nm}{N}\right]R^{m}Q^{n}
\end{equation}
In our discrete model the non-commutativity is determined by $N$.
As $N\rightarrow\infty$ the relation-operators $Q$ and $R$ commute.
However it would be hasty to conclude that as $N\rightarrow\infty$
we reach the continuum. The transition from the discrete to the continuum
is a subtle affair and many options are available. Let us define
\begin{equation}
L=Na\qquad p=\frac{2\pi}{L}
\end{equation}
Then 
\begin{equation}
\exp\left[2\pi i\frac{1}{N}\right]=\exp\left[ipa\right].
\end{equation}
What counts is the size of the available phase space and we may use
Planck's constant $h$ as a unit measuring the number of phase space
cells. Using rather $\exp\left[\frac{i}{\hbar}pa\right]$, equ.(\ref{eq:42})
becomes
\begin{equation}
QR=\exp\left[\frac{i}{\hbar}pa\right]RQ\label{eq:46}
\end{equation}
Approaching the continuum we may replace the discrete operators by
exponential forms 
\begin{equation}
R=\exp\left[\frac{i}{\hbar}pX\right]\label{eq:47}
\end{equation}
\begin{equation}
Q=\exp\left[\frac{i}{\hbar}aP\right].\label{eq:48}
\end{equation}
With $R$ and $Q$ unitary operators, $X$ and $P$ are hermitian
operators. From equs. (\ref{eq:46}), (\ref{eq:47}), (\ref{eq:48}),
we deduce
\begin{equation}
\left[X,P\right]=i\hbar.
\end{equation}
The foundational non-commutative law of Quantum Mechanics testifies
that there is a limit size $\hbar\sim pa$ in dividing the phase space.
With $p\sim mv\simeq mc$ we understand that $a$ represents the Compton
wavelength.

\section*{Conclusions}

\qquad{}We are used first to wonder about particles or states and
then about their interactions. First to ask about \textquotedblleft{}what
is it\textquotedblright{} and afterwards \textquotedblleft{}how is
it\textquotedblright{}. On the other hand, quantum mechanics displays
a highly relational nature. We are led to reorient our thinking and
consider that things have no meaning in themselves, and that only
the correlations between them are \textquotedblleft{}real\textquotedblright{}
\cite{key-11}. We adopted the Peircean relational logic as a consistent
framework to prime correlations and gain new insights into these theories.
The logic of relations leads us naturally to the fundamental quantum
rule, the probability as the square of an amplitude. The study of
a simple discrete model, once extended to the continuum, reveals that
only finite degrees of freedom can live in a given phase space. The
\textquotedblleft{}granularity\textquotedblright{} of phase space
(how many cells reside within a given phase space) is determined by
Planck\textquoteright{}s constant $h$. 

\qquad{}Discerning the foundations of a theory is not simply a curiosity.
It is a quest for the internal architecture of the theory, offering
a better comprehension of the entire theoretical construction and
favoring the study of more complex issues. We have indicated elsewhere
\cite{key-12} that a relation may be represented by a spinor. The
Cartan \textendash{} Penrose argument \cite{key-13,key-14}, connecting
spinor to geometry, allowed us to study geometries using spinors.
Furthermore we have shown that space-time may emerge as the outcome
of quantum entanglement \cite{key-15}. 

\qquad{}It isn\textquoteright{}t inappropriate to connect category
theory and relational logic, the conceptual foundations of quantum
mechanics, to broader philosophical interrogations. Relational and
categorical principles have been presented by Aristotle, Leibniz,
Kant, Peirce, among others. Relational ontology is one of the cornerstones
of Christian theology, advocated consistently by the Fathers (notably
by Saint Gregory Palamas). We should view then science as a \textquotedblleft{}laboratory
philosophy\textquotedblright{} and always link the meaning of concepts
to their operational or practical consequences.

\section*{Acknowledgment}

This work has been supported by the Templeton Foundation.

\end{document}